\documentclass[12pt,preprint]{aastex}

\usepackage{epsfig}
\usepackage{lscape,graphicx,natbib,helvet,graphicx,vpage,wrapfig,multicol,fancyhdr,epsfig,wrapfig}

\shorttitle{Deconstruction of active region AR10961}
\shortauthors{Noglik et al.}

\begin{document}
\title{Deconstructing active region AR10961 using STEREO, Hinode, TRACE \& SOHO}   
\author{Jane B. Noglik and Robert W. Walsh}   
\affil{Centre for Astrophysics, University of Central Lancashire, Preston, PR1 2HE}    
\author{Rhona C. Maclean}
\affil{Institute of Mathematics, University of St Andrews, The North Haugh, St Andrews, Fife, KY16 9SS}

\begin{abstract} 

Active region 10961 was observed over a five day period (2007 July 2-6) by instrumentation on-board STEREO, Hinode, TRACE and SOHO. As it progressed from Sun centre to the solar limb a comprehensive analysis of the EUV, X-ray and magnetic field data reveals clearly observable changes in the global nature of the region.

Temperature analyses undertaken using STEREO EUVI double filter ratios and XRT single and combined filter ratios demonstrate an overall cooling of the region from between 1.6 - 3.0~MK to 1.0 - 2.0~MK over the five days. Similarly, Hinode EIS density measurements show a corresponding increase in density of 27~$\%$.

Moss, cool (1~MK) outer loop areas and hotter core loop regions were examined and compared with potential magnetic field extrapolations from SOHO MDI magnetogram data. In particular it was found that the potential field model was able to predict the structure of the hotter X-ray loops and that the larger cool loops seen in 171~{\AA} images appeared to follow the separatrix surfaces. The reasons behind the high density moss regions only observed on one side of the active region are examined further.

\end{abstract}


\section{Introduction}

Currently the solar physics community has a fleet of dedicated satellites that are allowing us to probe the solar atmosphere over an extensive wavelength range. Since its launch in 1995, the Solar and Heliospheric Observatory (SOHO) has revolutionised our appreciation of the importance of magnetic connectivity from the photosphere through to the corona. High spatial resolution ($\sim$0.5$^{\prime\prime}$) EUV images from TRACE have altered our understanding of the dynamic nature of the corona. Subsequently, the launch of Hinode provides a high-resolution, EUV imaging spectrometer (EIS) and X-ray imaging telescope (XRT), along with the largest solar optical telescope (SOT) ever flown in space. The advent of such a high-resolution imaging spectrometer led directly to the production of coronal density maps (Young 2007); and more recently using XRT data, coronal temperature maps (Narukage et al. 2007). STEREO/EUVI on the other hand offers full disc solar images at four EUV wavelengths with a spatial resolution of 3$^{\prime\prime}$. STEREO also has the unique advantage that it consists of two identical satellites viewing the same coronal features at exactly the same time from different locations in space.

\begin{figure*}
\centering
\figurenum{1}
\epsscale{2.0}
\plotfiddle{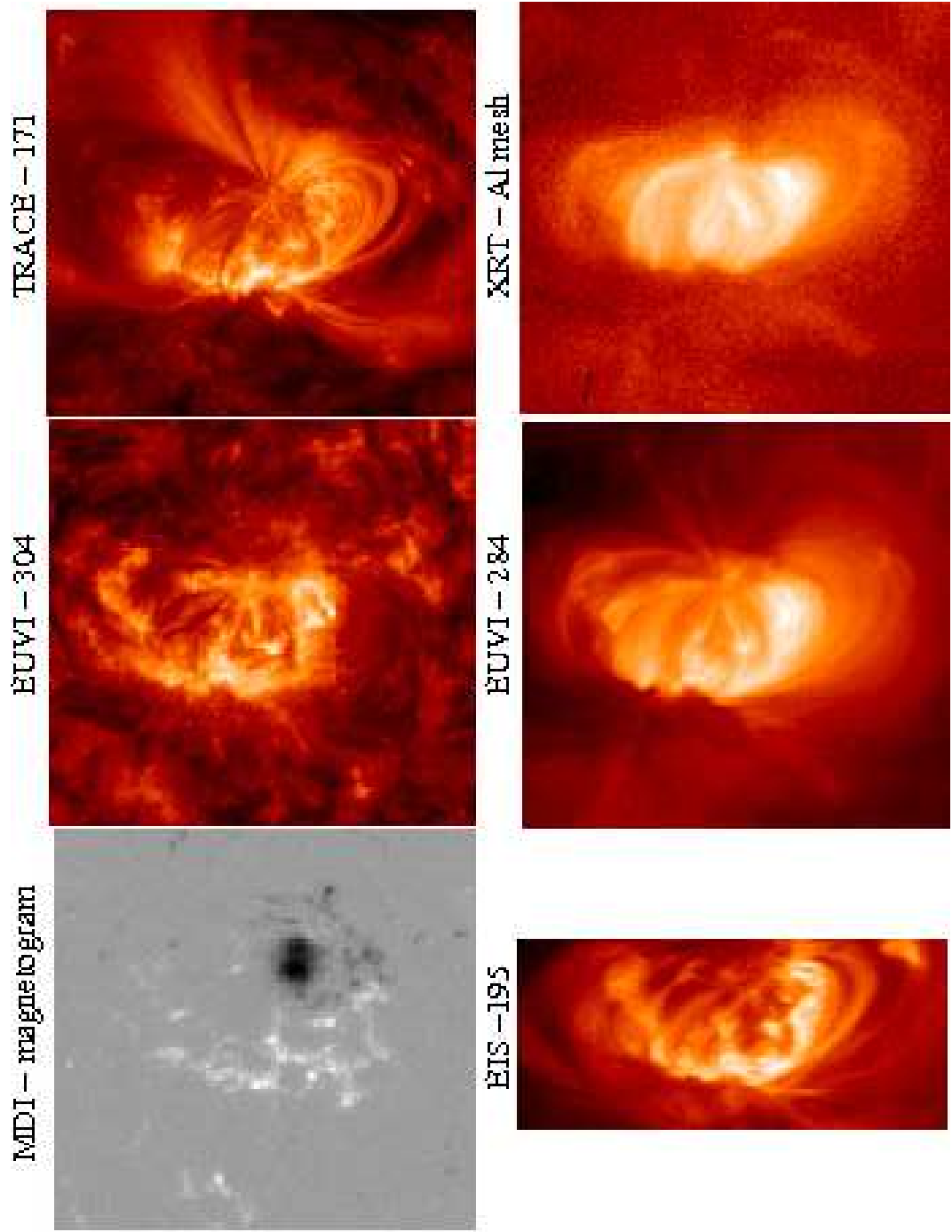}{1.in}{90.}{525.}{400.}{-20}{2000}
\caption{Active region images taken on Day 2 of the observations from all five instruments discussed in this paper. The images are approximately 250$^{\prime\prime}$ $\times$ 250$^{\prime\prime}$ apart from the EIS image which is $\sim$ 120$^{\prime\prime}$ $\times$ 250$^{\prime\prime}$. {\it Top:} From left to right; a SOHO MDI full-disc magnetogram, a STEREO~A EUVI 304~{\AA} image and a TRACE 171~{\AA} image. {\it Bottom:} From left to right; a Hinode EIS 195.12~{\AA} spectroscopic image, a STEREO~A EUVI 284~{\AA} image and an Al-mesh Hinode XRT image. \label{fig1}}
\end{figure*}

In this paper, we analyse data taken from five separate instruments upon these four different satellites over an observing period of five days. Our target was a typically small evolving active region AR10961; Figure~\ref{fig1} presents a close-up image of the active region as seen by each of the five instruments. The aim here is to use this unique coverage in order to understand the global change of this specific active region over the five-day period. By deconstructing the active region into its temperature, density, velocity and magnetic field structures, we can examine the relationship between the photospheric and the coronal emission (both X-ray and EUV) as well as the origin of high/low temperatures and densities associated with individual features.

Methods for determining physical properties, such as temperature (T($s$)) and density ($\rho$($s$)), have been highly debated over the last decade. It has been shown that the application of a correct background subtraction to EUV imaging data taken with SOHO and TRACE is essential in order to obtain true temperature diagnostics (Del Zanna \& Mason 2003). Even in light of this, previous attempts to discover temperatures of coronal structures have proven to be inconclusive or even contradictory (Schmelz et al. 2003, Aschwanden et al. 2005, Cirtain et al. 2007, Noglik et al. 2007). 

Tripathi et al. (2008) analysed EIS density maps over a wide range of temperatures, for an AR observed near disc centre on 2007 May 1. They found that moss regions were primarily focused to one side of the AR and that the electron densities of the moss regions decreased with increasing temperature. They also found that these moss regions mapped to the positive polarity observed on the MDI magnetograms but there were no moss regions over the negative polarity where a large sunspot was visible. 

Del Zanna (2008) also used Hinode EIS data to investigate loop structures within an AR. When looking at the flows present within these loops he found persistent redshifts in the cooler lines in most loop structures. He also saw blueshifts of $\sim$5 - 20~km~s$^{-1}$ in Fe~{\sc xii} (higher velocities in hotter lines) present in areas of weak emission and in areas in between ``hot'' loops ($>$2~MK) and ``cooler'' ($\sim$1~MK) loops.

In the following, Section 2 describes briefly the observational datasets, whilst Section 3 begins to investigate the STEREO~A EUVI data. Section 4 utilises the XRT data for global and structural temperature analysis, and Section 5 analyses the EIS density and velocity maps over the observational period. Section 6 shows the initial MDI magnetic field extrapolation with comparison to EUV and X-ray data. Section 7 outlines the discussion and conclusions.

\section{Observations: an overview of the active region}

\begin{figure*}
\centering
\figurenum{2}
\plotfiddle{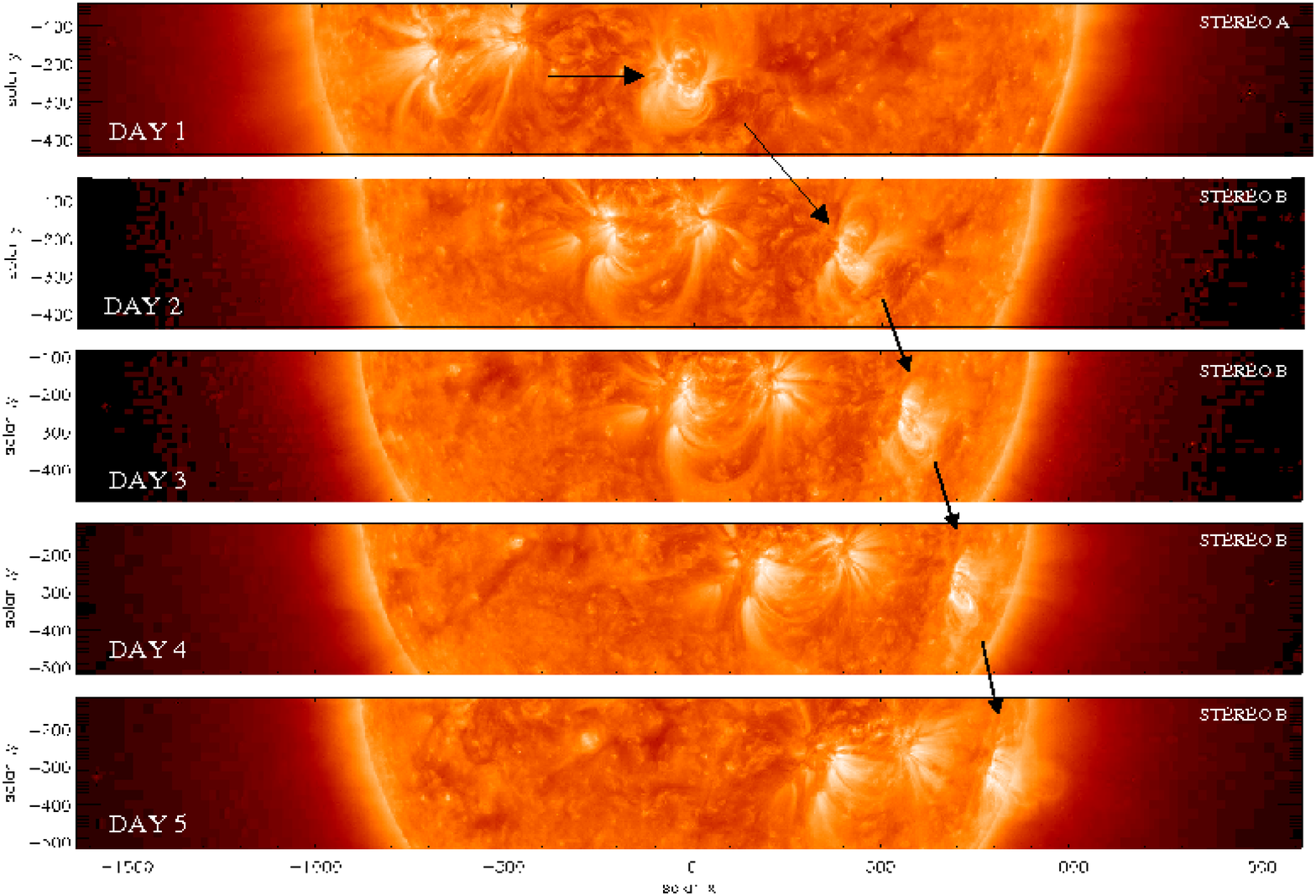}{1.in}{90.}{550.}{390.}{0}{2000}
\caption{Evolution of the active region over the five day observing period. The data were taken from both STEREO~A \& B at 171~{\AA}. The black arrows point to the active region investigated in this paper. \label{fig2}}
\end{figure*}

\begin{figure*}
\centering
\figurenum{3}
\plotfiddle{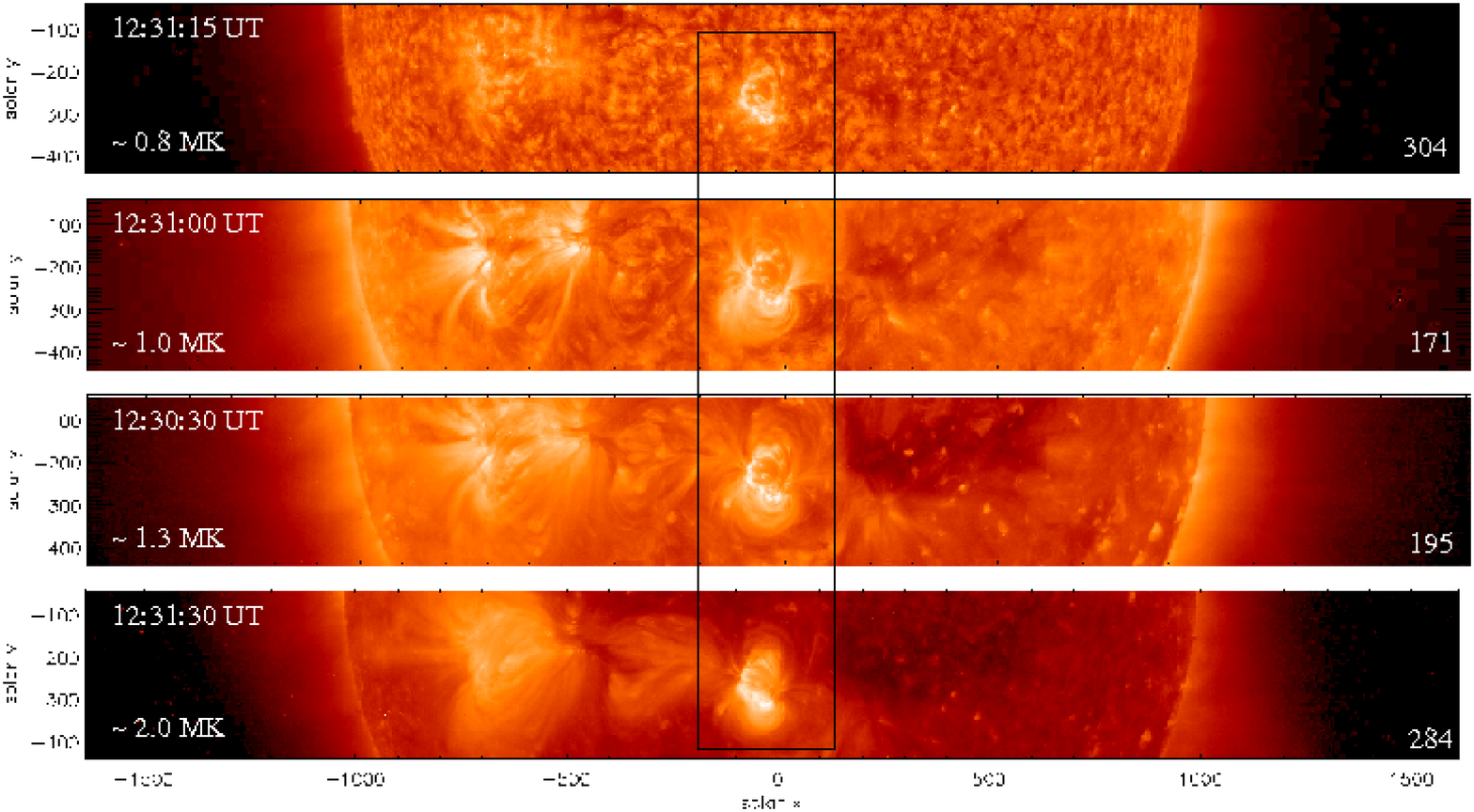}{1.in}{90.}{550.}{370.}{0}{2000}
\caption{STEREO~A EUVI images taken on Day 1 of the observing run at 304~{\AA}, 171~{\AA}, 195~{\AA} and 284~{\AA}, respectively.\label{fig3}}
\end{figure*}

The data analysed in this paper were taken as part of the Hinode Observing Programme (HOP) 18. It captured a small active region which was situated near the disc centre on 2007 July 02 and followed its progression to the solar limb over a period of five days. Generally, STEREO EUVI data were taken consistently throughout the five day observing run, though note that, there were few STEREO~A datasets taken on Day 4 and no STEREO~B data on Day 1. Figure~\ref{fig2} shows the evolution of the active region as seen by STEREO EUVI at 171~{\AA} while Figure~\ref{fig3} displays the active region on Day 1 at four differing EUV temperatures. Figure~\ref{fig3} shows that at the coolest temperature of $\sim$0.8~MK (304~{\AA}) strong moss emission is seen to dominate the left hand-side of the AR, but no other features are easily discernible. Moving up to $\sim$1~MK (171~{\AA}) immediately large loops are seen to emanate from the central eastern and western regions, traversing above and below the AR core; at the same time, the intense moss regions to the left are still visible. At $\sim$1.3~MK (195~{\AA}) these long extended cool loops begin to fade but the moss regions remain bright and hints of large structures connecting this AR to a larger AR further to the east are now apparent. At temperatures of 2~MK or above (284~{\AA}) the central core of the AR starts to brighten in its intensity, clearly showing loop structures. The cooler 1~MK loops have disappeared and hot loops connecting this AR to a trailing larger AR are also seen clearly. 

For the first two days, XRT observed only in the Al-mesh and Ti-poly filters. From Day 3, however, runs using eight different filters were undertaken with a cadence of a few seconds. The EIS observations were taken using the 1$^{\prime\prime}$ slit with an exposure time of 10~s starting on Day 2 of the observations. It took $\sim$23~mins to raster across this active region. However, as the five day observational run progresses and the AR heads towards the limb, the apparent ``area'' occupied by the AR decreases and therefore, the total time for EIS to raster across the region is reduced. There are two sets of density-sensitive pairs available within this EIS dataset, the Fe~{\sc xii} 195.12 / 186.88~\AA\ pair and the Fe~{\sc xiii} 202.04 / 196.54~\AA\ pair. This EIS raster was run once on Day 2, and then three times every day thereafter. TRACE followed the progression of the active region for the full five days in the 171~{\AA} passband only, but TRACE still has the highest spatial resolution (0.5$^{\prime\prime}$) of all the solar EUV imagers. 

\section{STEREO~A EUVI data analysis}

\begin{figure*}[t]
\centering
\figurenum{4}
\plotfiddle{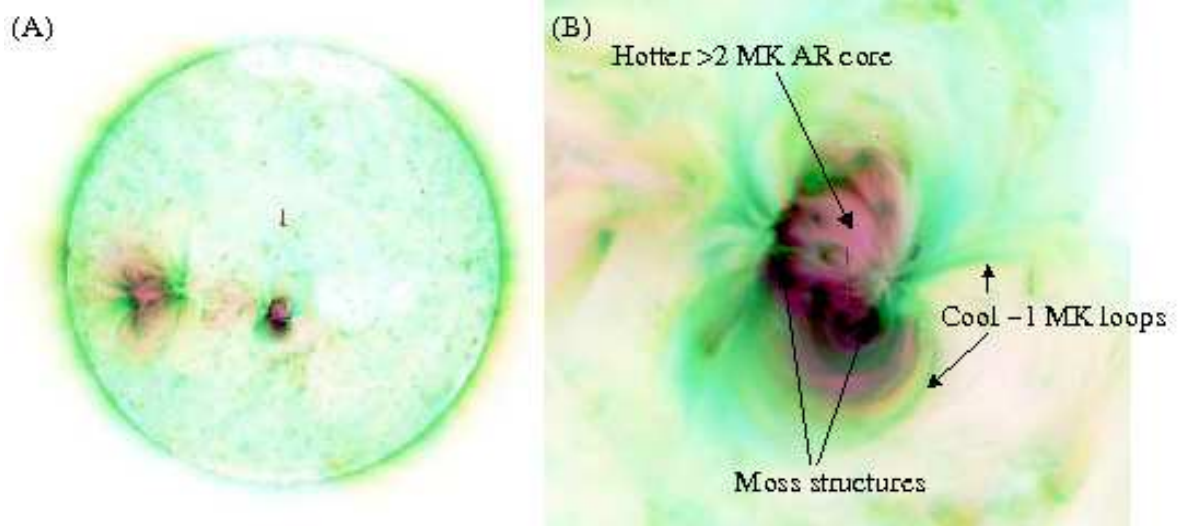}{1.in}{0.}{450.}{200.}{0}{0}
\caption{(A) Full disc composition image using EUVI imagers from STEREO A, at 171~\AA\ (blue), 195~\AA\ (yellow) and 284~\AA\ (red). (B) Close-up of the active region taken from image (A).\label{fig4}}
\end{figure*}

Figure~\ref{fig4} is a composite image made using three different wavelengths from STEREO~A taken on Day 1. The image was constructed using the software package Aladin v5.0 (Bonnarel et al. 2000). Here the 171~{\AA} (1~MK) emission is shown in blue, the 195~{\AA} (1.3~MK) emission in yellow and the 284~{\AA} ($\sim$2~MK) emission in red. On the full disc image (A) we can see that the majority of the solar emission is blue or green, the green colour is produced when the 171~{\AA} (blue) emission and the 195~{\AA} emission (yellow) are overlapping. The only hotter or redder regions are clearly seen within the two active regions that are present and along the loop structures that are connecting the two regions. What we see from the close-up image Figure~\ref{fig4}~(B), is that the centre of the AR appears dominated by the hotter, in this instance redder, emission from 284~{\AA}, and then moving away from the AR core it becomes greener/bluer (cooler) and eventually the blue colours start to become evident, showing the coolest emission at the edges of the AR. Also, the moss regions are black, demonstrating that the moss is very bright in all three of these wavelengths and is therefore emitting across a fairly wide temperature range (including 304~{\AA} as shown in Figure~\ref{fig3}). If the 284~{\AA} image is replaced in the composite (B) by the cool 304~{\AA} image then the centre of the AR remains mostly white with very little emission present, as the 304~{\AA} emission only displays very high intensities over the moss regions. 

\begin{figure*}[h!]
\centering
\figurenum{5}
\plotfiddle{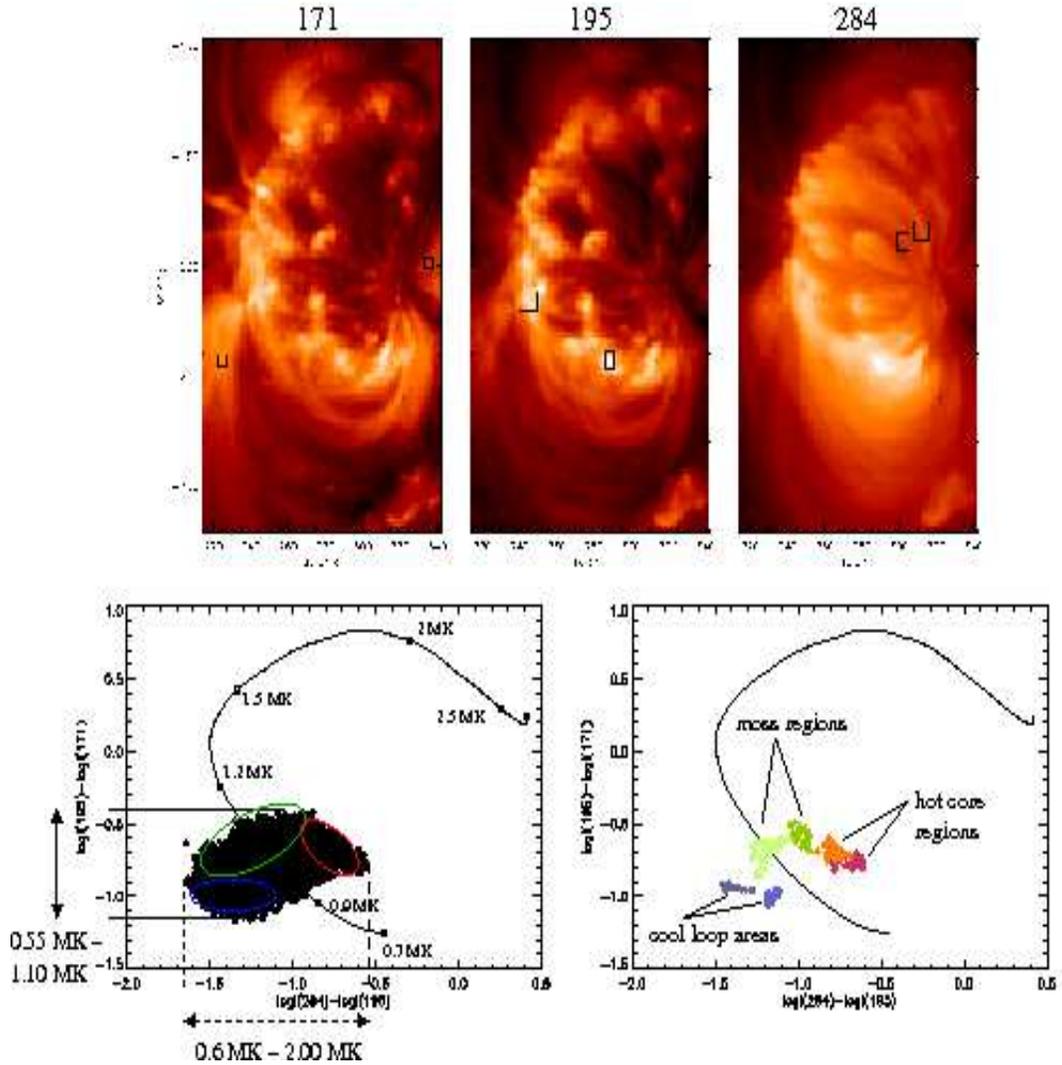}{1.in}{0}{400}{400}{-0}{-0}
\caption{{\it Top:} Three STEREO~A EUVI images taken from Day 2, at 171~{\AA}, 195~{\AA} and 284~{\AA}. The boxes on these images show the two cool loop patches (171~{\AA}), the two moss regions (195~{\AA}) and the two hot core regions (284~{ \AA}) used for the analysis. {\it Bottom left:} Colour-colour curve of the 195~{\AA}/171~{\AA} ratio against the 284~{\AA}/195~{\AA} ratio, predicted from the EUVI~A temperature response functions. The stars are the actual ratios for each pixel on the above images. {\it Bottom right:} Colour-colour curve this time only showing the intensity ratios for the areas within the boxes marked on the above images. Blue points for the cool 1~MK loop areas, green points for the moss regions and red points for the hotter active region core.\label{fig5}}
\end{figure*}

Figure~\ref{fig5} ({\it top}) shows a triplet dataset (171~{\AA}, 195~{\AA} and 284~{\AA}) taken by STEREO~A on Day 2 of the observations. On the bottom of Figure~\ref{fig5} is the colour-colour curve for the EUVI instrument on-board STEREO~A, created using predicted temperature response functions for each wavelength, where every point on the curve is related to a specific temperature value. This method has been called the double filter ratio temperature analysis technique (see Noglik \& Walsh 2007; Chae et al. 2002). The points plotted across the left colour-colour diagram represent the actual intensity ratios for each pixel in the above images. The points spread across the lower half of the diagram, concentrated towards the cooler temperatures of 0.9~MK - 1.2~MK. Also, note that the range of temperatures viewed by the 284~{\AA}/195~{\AA} ratio only (0.60 - 2.0~MK) appears larger than the temperatures seen by the single 195~{\AA}/171~{\AA} ratio (0.55 - 1.1~MK). Looking at the 284~{\AA} EUVI data we know that there is definitely emission present within the AR at 2~MK or above. Therefore, the fact that the maximum temperature deduced from the 195~{\AA}/171~{\AA} single filter ratio is 1.1~MK suggests that this ratio is not very sensitive to temperatures above $\sim$1~MK. 

The three images each display two small boxes, the boxes on the 171~{\AA} image contain patches of cooler loop emission, on the 195~{\AA} image the boxes contain regions of moss and on the 284~{\AA} image the boxes cover parts of the AR core. Only the intensity ratios from inside all of these specific boxes are plotted on the colour-colour curve shown on Figure~\ref{fig5} bottom right, this allows us to compare the different areas within the AR. The 171~{\AA} loop patches are plotted in blue, the moss regions in green and the core AR areas in red; these ratios are also highlighted using the same colours on the left hand diagram. It is clear that the intensity ratios for each of these three different structures occupy separate areas on the colour-colour diagram, and that these areas are not seen to overlap. This suggests that there is an identifiable temperature difference between the three regions that can be observed with EUVI data. It is seen that the ratios from the moss regions lie the closest to the colour-colour curve; this could be due to the fact that these are the only areas observed in all three wavelengths. This can be compared to the fact that the AR core emission is missing from the 171~{\AA} images and the cooler loop emission is missing from the 284~{\AA} images. The cool/cooling loop patches have moved away from the curve towards the bottom left hand corner, demonstrating a decrease in both the filter ratio values i.e. the 195~{\AA} intensity becomes stronger in comparison to the 284~{\AA} intensity as well as the 171~{\AA} intensity becoming stronger when compared to the 195~{\AA} intensity. Hence, a cooler temperature is dominating in these regions which is what would be expected. The exact opposite is true for the hotter loop areas as they move to the top right of the diagram, therefore demonstrating a movement into the filters with higher peak temperatures. The observed movement of these points away from the colour-colour curve is likely to be due to the dynamic heating/cooling of the structures or the absence of emission in one of the passbands or a combination of both.  


\section{Hinode / XRT temperature analysis}


XRT data is available for this AR over the entire observing period. However, during Day~1 and 2 of the observations only two different filters were used; Al~mesh and Ti~poly. Starting on Day~3 of the observations multi-filter data were recorded which included, Al~poly, C~poly, Be~med, Be~thin, Ti~poly, Al~thick, Al~poly/Ti~poly and C~poly/Al~thick.

\begin{figure*}
\centering
\figurenum{6}
\plotfiddle{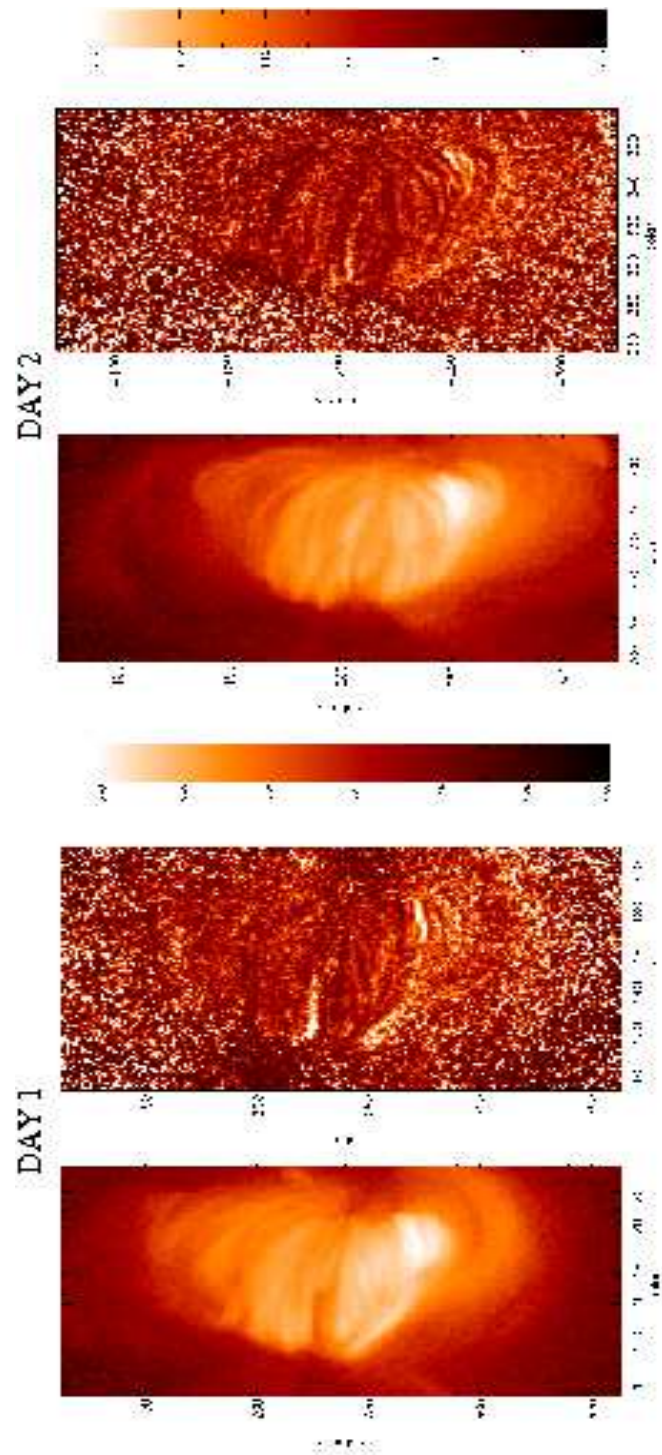}{1.in}{90}{550}{250}{-0}{2000}
\caption{An Al~mesh intensity image and a single filter ratio temperature map constructed for Day 1 \& 2, between 12:00~UT and 13:00~UT. \label{fig6}}
\end{figure*}

Using data taken between 12:00~UT and 13:00~UT, Figure~\ref{fig6} shows Al~mesh intensity images on a logarithmic scale from Day 1 and 2, along with a corresponding temperature map for each day. The temperature maps were constructed using the predicted temperature response functions for the Al~mesh and Ti~poly filters, obtainable with {\it calc\_xrt\_temp\_resp.pro} in sswidl. It is possible to create such maps using these predicted response functions as there is generally only one temperature solution for any given ratio. The temperature range of the active region was found to be 1~MK to 3~MK with the majority of the emission lying above 2~MK. This could explain the difference in appearance of the AR when seen at EUV wavelengths and at X-ray wavelengths, as the EUV emission lies between 0.5~MK and 2.0~MK and the majority of the X-ray ``loop'' emission is seen to be above 2.0~MK. For a daily comparison of the XRT data to EUVI 284~{\AA} image see Figure~\ref{fig9} (discussed further below). 

\begin{figure*}
\centering
\figurenum{7}
\plotfiddle{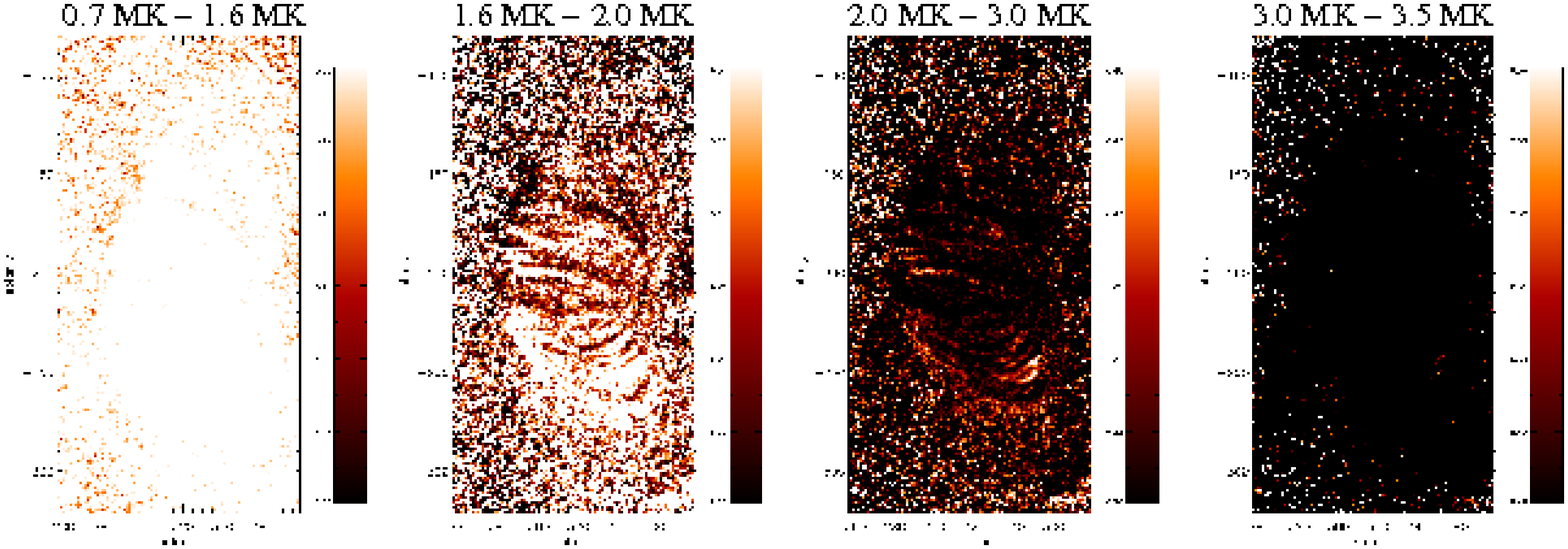}{1.in}{90}{550}{250}{-0}{2000}
\caption{Four temperature maps from Day 2 (using the same data shown in Figure~\ref{fig6}) each depicting a different temperature range to indicate where the majority of the AR emission lies. \label{fig7}}
\end{figure*}

Figure~\ref{fig7} shows a filtering of the temperature map for Day~2. This demonstrates that the XRT data does not show AR emission below 1.6~MK. Examining the slightly higher temperature range of 1.6~MK - 2.0~MK, results in the AR emission from loop structures becoming evident, as well as emission towards the north of the AR. Moving up to 2.0~MK - 3.0~MK, some bright loop structures in the central and southern AR areas are now visible. The final map shows that above 3.0~MK, again there is negligible emission from the AR.

\begin{figure*}
\centering
\figurenum{8}
\plotfiddle{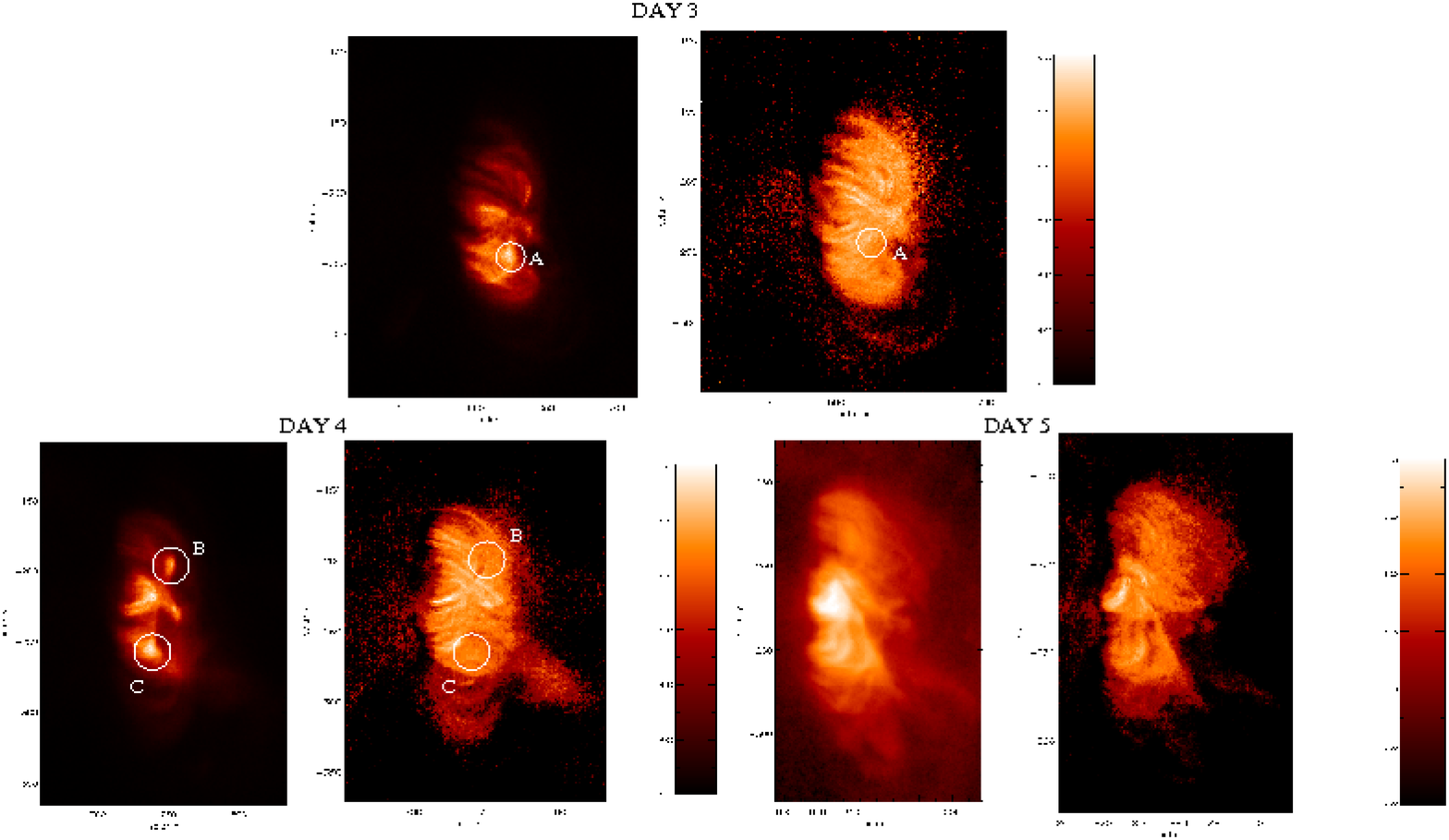}{1.in}{90}{550}{350}{-0}{-2000}
\caption{An Al~poly intensity image and a CIFR temperature map calculated for Days 3, 4 \& 5 using data taken between 12:00~UT and 13:00~UT on each day. \label{fig8}}
\end{figure*}

\begin{figure*}
\centering
\figurenum{9}
\plotfiddle{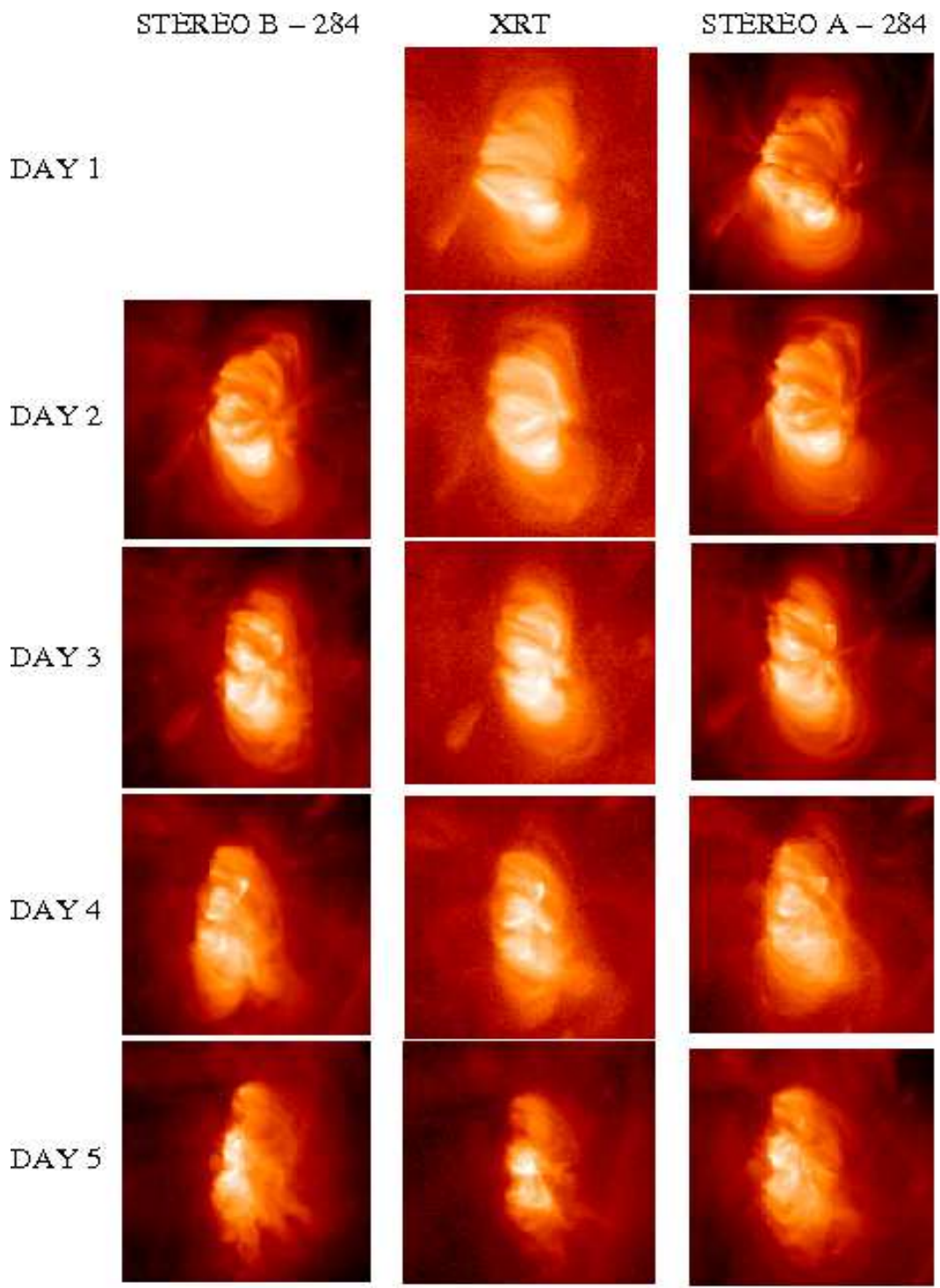}{1.in}{0}{400}{550}{-0}{-0}
\caption{{\it Left:} STEREO~B EUVI 284~{\AA} images taken at $\sim$ 12:30~UT for each day (apart from Day 1) of the observing run. {\it Middle:} Hinode XRT images taken at approximately the same time as the EUVI images.{\it Right:} STEREO~A EUVI 284~{\AA} images taken at $\sim$ 12:30~UT for each day of the observing run. \label{fig9}}
\end{figure*}

Using only two XRT filters to produce these maps it is quite difficult to see and locate exact features within the temperature diagrams that directly relate to the intensity images. In contrast, Figure~\ref{fig8} represents Al~poly images taken on Days 3, 4 and 5, alongside temperature maps for each day constructed using at least five different filters. We have not taken the logarithm of the intensity images on Days 3 \& 5 (unlike the other intensity images), because the distinctive features within the central active region were only clearly visible on a linear scale. These temperature maps were created using the combined improved filter ratio (CIFR) technique (see Reale et al. (2008) for full details of this method). The method uses the ratio between the geometric mean of the emission detected in all the available filters and the emission from a single filter. However, not all eight of the observed filters were always used as a few of the observations were over-exposed whilst some others were under-exposed leading to low count rates. 


The temperature maps displayed in Figure~\ref{fig8} show much more intricate detail than the ones created from Day 1 \& 2. Temperature features within the active region core are easily identifiable. Hotter ($\sim$2~MK) and cooler ($\sim$1.5~MK) loops are observed sitting along side each other and generally the loop tops appear around 40$\%$ hotter (2~MK compared with 1.4~MK) than the legs of the loop structures. The hotter loops do not always correspond to the brighter emission loops, for instance, on Days 3 \& 4 the bright emission towards the south (regions A \& C on Figure~\ref{fig8}) produces a very average global temperature of $\sim$1.66~MK. Also on Day 4, the intense brightening in the north (region B on Figure~\ref{fig8}) is actually seen to be dark, and therefore cooler on the temperature map. This indicates that the enhanced emission is a density effect which will be investigated using EIS data in the next section. It is important to note that the temperature scale only ranges from 1~MK to 2~MK; whereas previously on Days 1 \& 2 the majority of the emission was seen to be above 2~MK. Generally, these temperature scales seem low for X-ray images, and in comparison to the 171~{\AA} and 195~{\AA} intensity images from EUVI there is very little similarity in the morphology of the AR. However, looking at the 284~{\AA} images from Days 3, 4 \& 5 there is a similarity in the structures seen between this passband and the X-ray wavelengths (Figure~\ref{fig9}). This is not the case on Days 1 \& 2 where the temperatures generally remained above 1.7~MK. 

Considering Figure~\ref{fig9}, on Days 1 \& 2 the EUVI 284~{\AA} images from STEREO A \& B do display some similarity to the corresponding XRT images. However, by Days 3, 4 \& 5, there are genuine one-to-one correspondences of bright patches observed in all three instruments. Thus, the temperature values calculated from XRT appear to reflect this, demonstrating a global decrease in the temperature over time. Obviously, when comparing XRT to EUVI data we need to be careful as Hinode and STEREO are viewing this AR from different angles. However, this observing run was carried out only seven months after the launch of STEREO when the two satellites were still very close to Earth (the separation angle from STEREO~A to Earth was $\sim$10 degrees and from STEREO~B to Earth only $\sim$6.3 degrees).  

\section{Hinode / EIS density analysis}

The EIS data was prepared for analysis using {\it eis\_prep}; {\it eis\_getwindata}; {\it eis\_auto\_fit} and {\it eis\_orbit\_correction}. The 186.88~\AA\ Fe~{\sc xii} line and the 196.54~\AA\ Fe~{\sc xiii} line cannot be simply extracted using a single Gaussian fit  (see Young et al. 2007). Therefore, separate routines written to isolate these particular lines were used instead of {\it eis\_auto\_fit}; the Gaussian fits can then all be checked using {\it eis\_fit\_viewer}. Also note, that there can often be a tilt present between the images in the y-direction which needs to be corrected for before performing the density analysis. The density maps were then created, for Fe~{\sc xii} and Fe~{\sc xiii}, using the extracted intensities along with the CHIANTI atomic database (Dere et al. 1997, Landi et al. 2006). On Day 2 the EIS observations were taken between 06:00 and 07:00~UT. For Days 3, 4 \& 5 three sets of observations were taken each day; here we have concentrated upon the dataset taken closest to 12:30~UT for comparison to the XRT data analysed in the previous section.

\begin{figure*}[!h]
\centering
\figurenum{10}
\plotfiddle{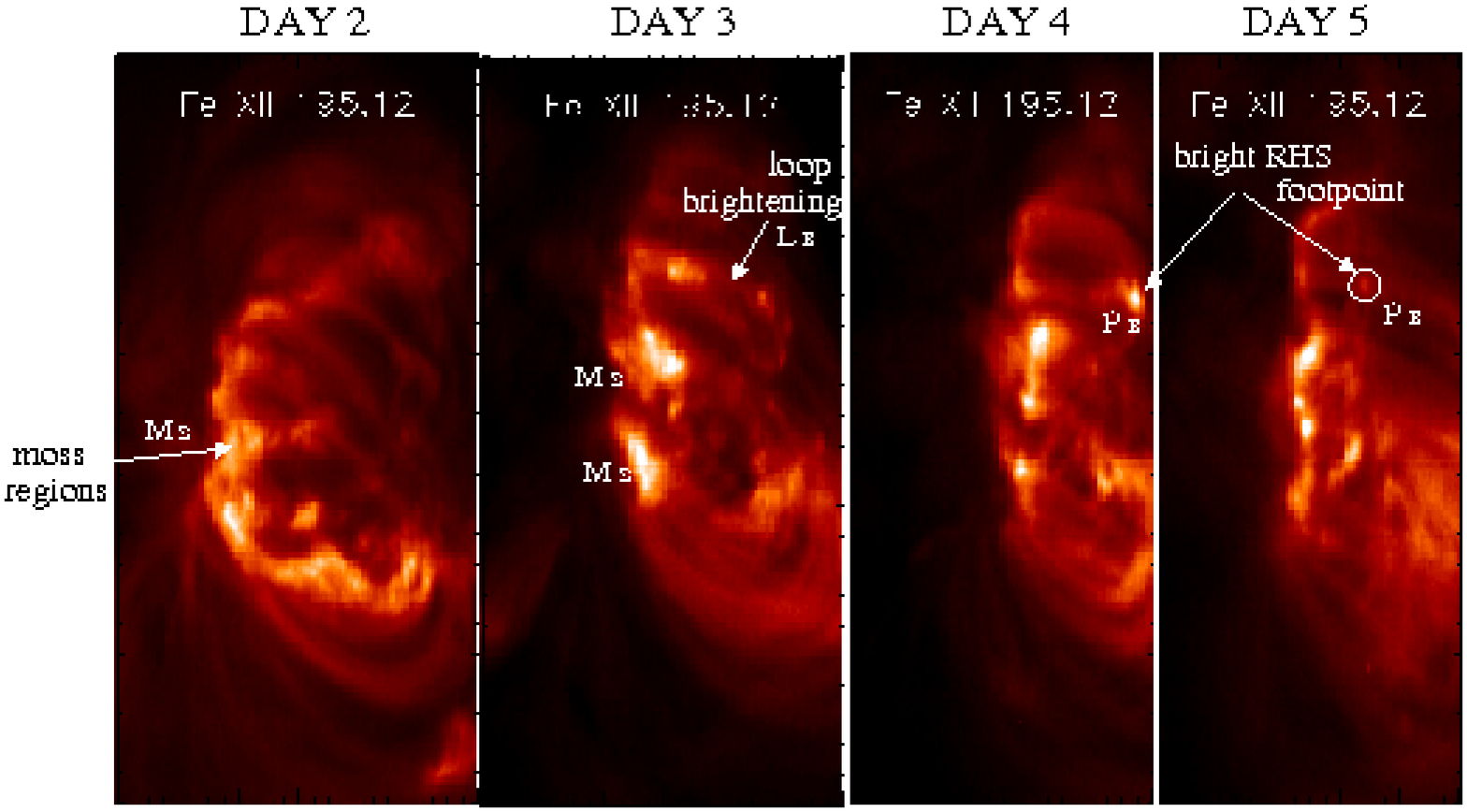}{1.in}{0}{350}{200}{-50}{-0}
\caption{Fe~{\sc xii} intensity images taken at $\sim$13:00~UT for Days 2, 3, 4 \& 5. \label{fig10}}
\end{figure*}


In Figure~\ref{fig10} the intensity images created from the Fe~{\sc xii} 195.12~{\AA} spectral line are displayed for Days 2 to 5 (there is no EIS data available for Day 1 of the HOP). The Fe~{\sc xiii} lines and the 186~{\AA} Fe~{\sc xii} line all show almost identical features and therefore only one set of intensity images has been shown. Note that, not surprisingly, these EIS images look very similar to the 195~{\AA} images taken with the EUVI instrument on STEREO. 

On Day 2 we see the intense ``mossy'' patches down the left-hand side and the base of the AR (indicated by an arrow and the letters M{\sc s}); as mentioned earlier this moss was also evident in the four EUV passbands viewed with STEREO. The right-hand side of the AR appears darker but some faint larger loops are visible to the south. 

On Day 3 we see two large (one central and one southern) bright moss regions (M{\sc s}), on the left hand side, that have become separated; and a loop feature (L{\sc b}) begins to brighten across the north of the AR (also indicated by arrows). On Days 4 \& 5 the AR moves around towards the limb of the Sun and unfortunately the tops of the loop structures have been missed from these observations. On Day 4, there are now some very bright footpoints visible (F{\sc b}) especially towards the top right-hand side, which is the first bright area to become visible on this side of the AR. Some hints of this brightening are possibly seen in connection with the loop structure that becomes visible on Day 3. By Day 5 the AR has rotated so far onto the limb that one cannot see clearly this footpoint (F{\sc b}) although an arrow indicates its location.   

\begin{figure*}
\centering
\figurenum{11}
\plotfiddle{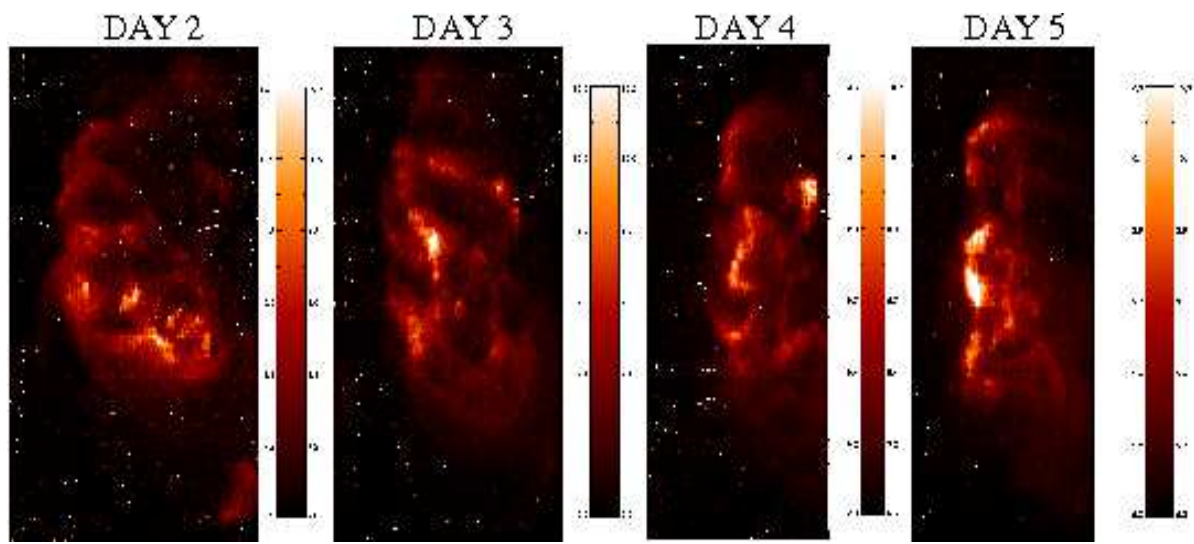}{1.in}{0}{450}{200}{-0}{-0}
\caption{Fe~{\sc xii} density maps corresponding to the intensity images shown in Figure~\ref{fig10}. \label{fig11}}
\end{figure*}


\begin{figure*}
\centering
\figurenum{12}
\plotfiddle{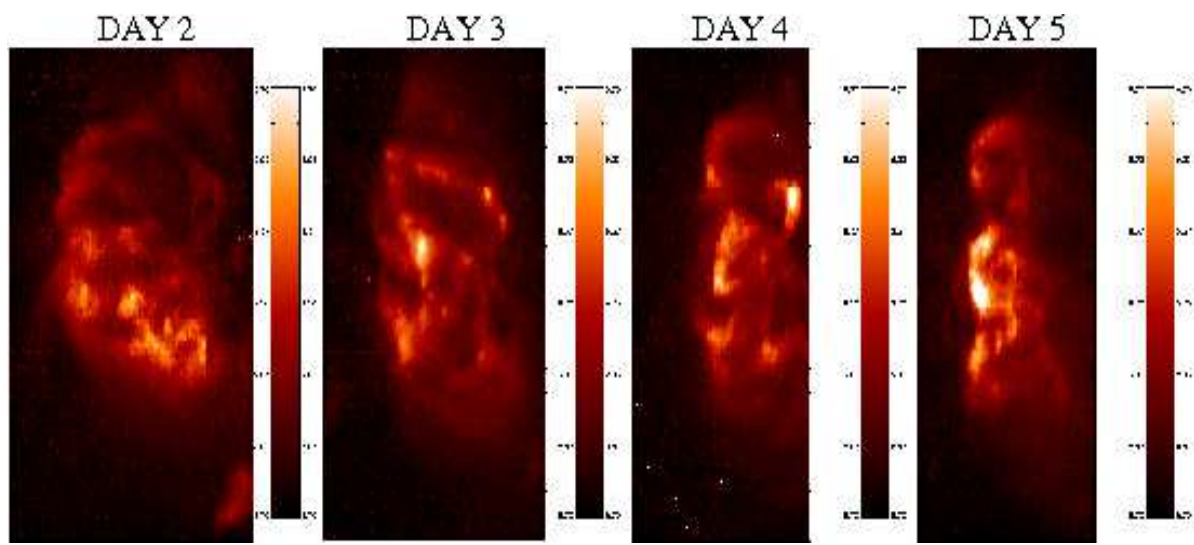}{1.in}{0}{450}{200}{-0}{-0}
\caption{Fe~{\sc xiii} density maps corresponding to data taken at the same time as the Fe~{\sc xii} intensity images shown in Figure~\ref{fig10}. \label{fig12}}
\end{figure*}


Figures~\ref{fig11} \& \ref{fig12} represent the densities calculated from the intensity images for Fe~{\sc xii} and Fe~{\sc xiii} respectively. The densities calculated from Fe~{\sc xii} are slightly higher than those calculated using Fe~{\sc xiii} as reported by Young et al. (2008). However, in both Fe~{\sc xii} and Fe~{\sc xiii} the density maps reflect larger densities in areas of higher intensity. The bright footpoint seen on the right-hand side of the AR on Day 4 also seems to reflect a large increase in density. This is the same bright feature which was seen in the XRT intensity images but which was represented by lower values on the temperature map; therefore validating the proposed density effect.

\begin{figure}[!h]
\centering
\figurenum{13}
\plotfiddle{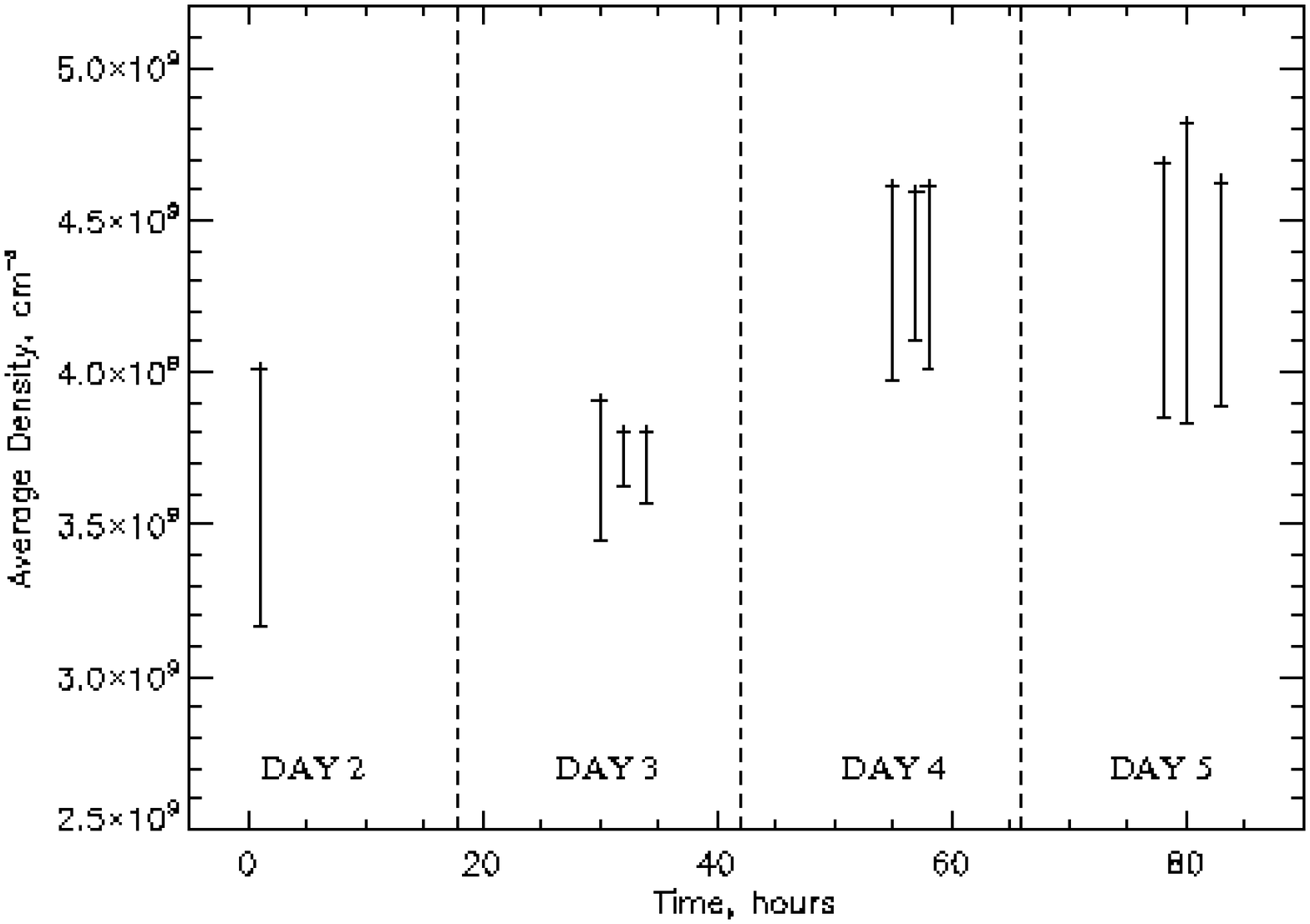}{1.in}{0}{300}{200}{-0}{-0}
\caption{Plot showing the average active region density over the four days that EIS spectroscopic data was available. \label{fig13}}
\end{figure}


Figure~\ref{fig13} shows the average calculated density (with a lower limit of 2.0$\times$10$^9$~cm$^{-3}$), as seen by Fe~{\sc xii}, over the entire active region against its observation time. There were three EIS runs taken on Days 3, 4, \& 5. Therefore, three points are plotted for each of these days. The average density remains very similar for observations taken on the same day. Overall, the average density of the active region appears to be increasing slightly every day, from $\sim$ 2.6$\times$10$^9$cm$^{-3}$ on Day 2 to 3.5$\times$10$^9$cm$^{-3}$ on Day 5 (an increase of $\sim$27~$\%$). In conclusion, the XRT observations from section 4 showed an apparent global decrease in temperature of the AR over time and now EIS observations seem to show an increase in density over the last four days of the observations.

\section{Hinode / EIS velocity maps}

\begin{figure*}
\centering
\figurenum{14}
\plotfiddle{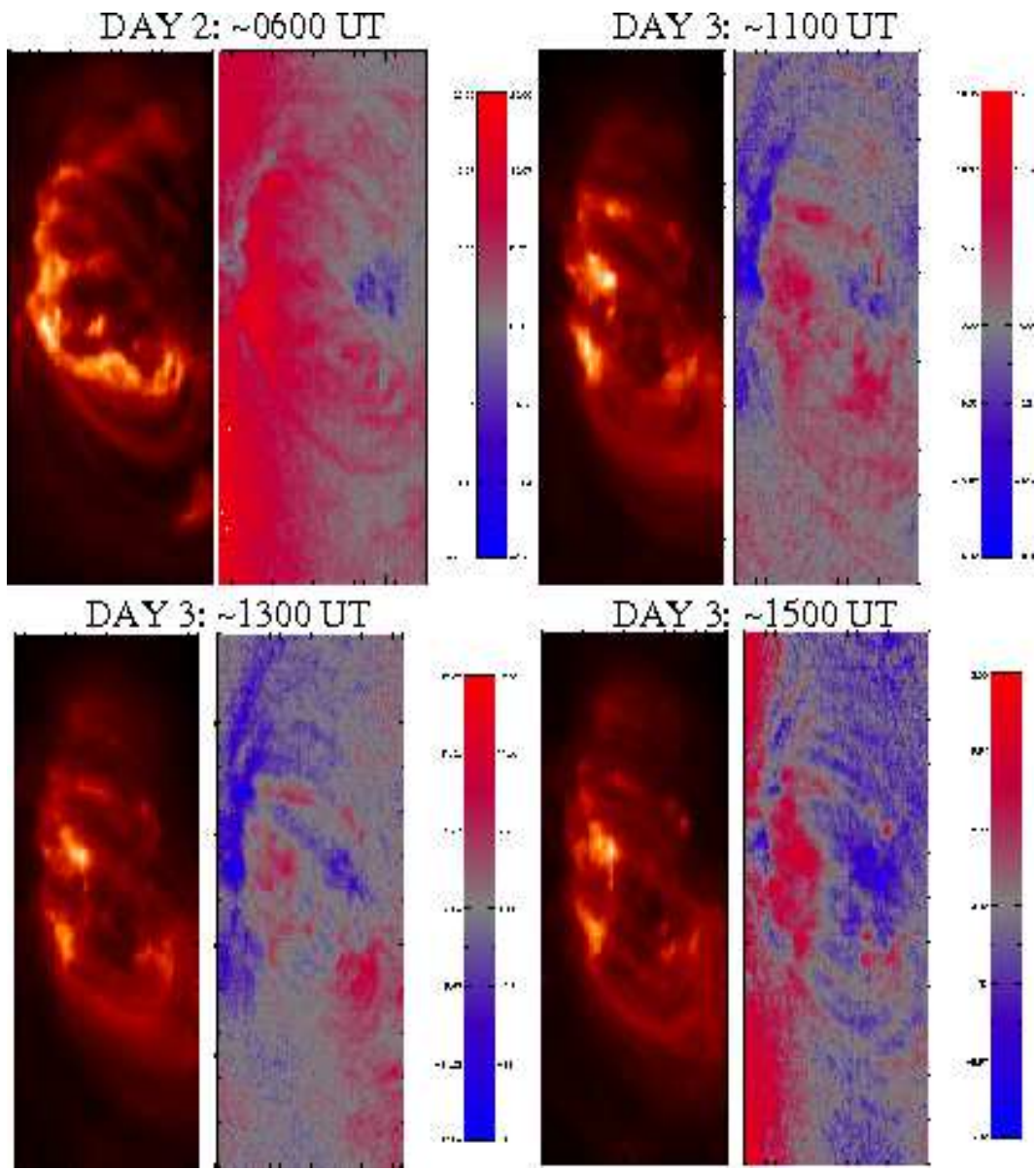}{1.in}{0}{400}{450}{-0}{-0}
\caption{Velocity maps from EIS data for Days 2 \& 3 presented alongside their intensity images for Fe~{\sc xii}. The strong intensities also represent the high density regions. \label{fig14}}
\end{figure*}

\begin{figure*}
\centering
\figurenum{15}
\plotfiddle{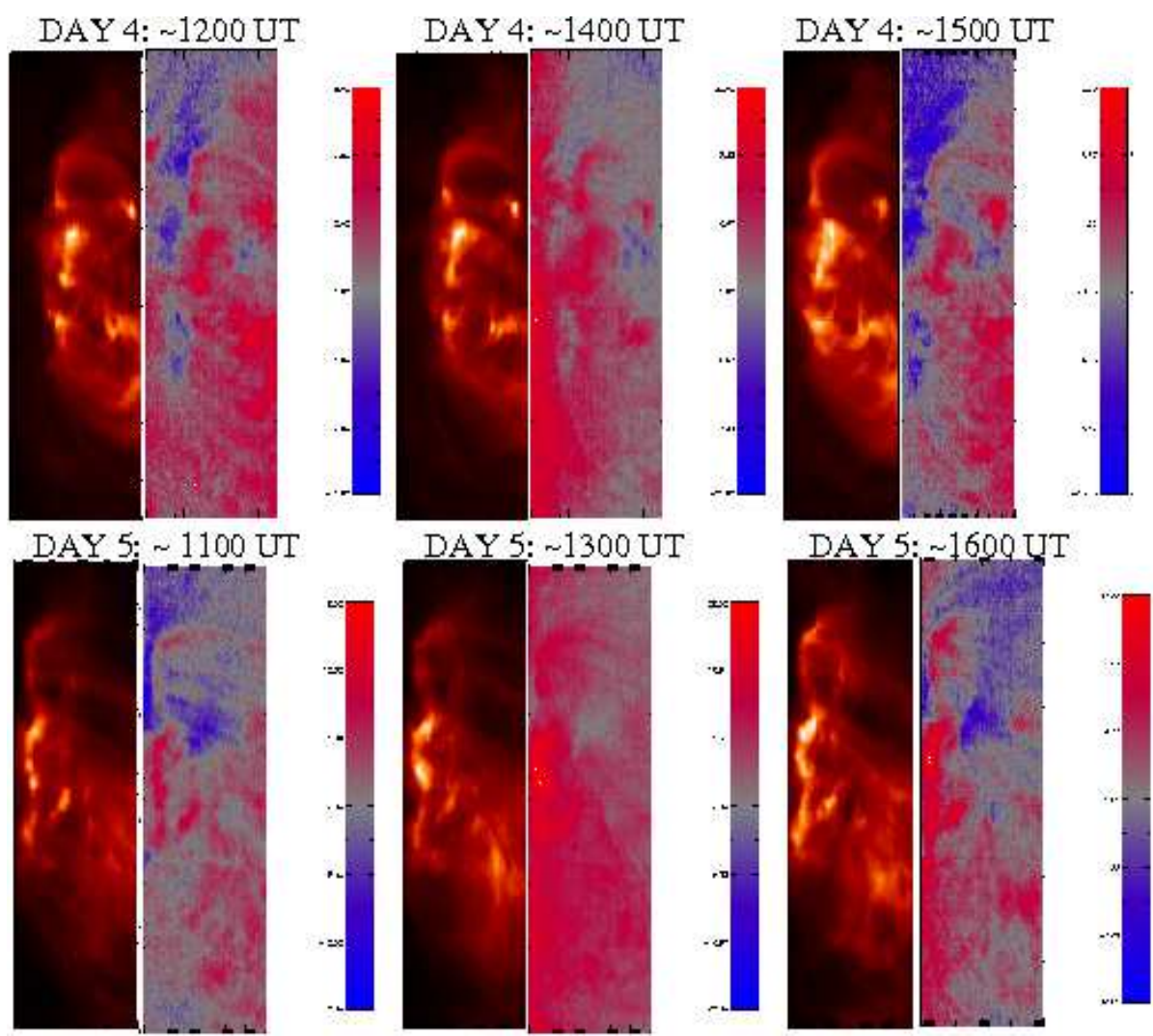}{1.in}{0}{450}{400}{-0}{-0}
\caption{Same as Figure~\ref{fig20} for Days 4 \& 5. \label{fig15}}
\end{figure*}

A brief look at the velocity maps, taken with EIS on Days 2 \& 3 (Figure~\ref{fig14}), shows blueshifts of $\sim$ 10 - 15~km~s$^{-1}$ over the right-hand side hot loop footpoints. These blueshifts are situated above the negative sunspot polarity. Gurman \& Athay (1983) as well as Marsch et al. (2004) reported seeing blueshifts above sunspots, but it was not clear how they related to the magnetic field. The only other blueshifts are seen in boundary sharp regions, between the hotter and cooler loop regions, and in low density regions, in exact correlation with Del Zanna (2008). These blueshifts are often thought to be the signature of heating/chromospheric evaporation during reconnection. The hot loop tops all exhibit fairly strong redshifts ranging from 13 - 25~km~s$^{-1}$ over the five-day observing period, as shown in Figures~\ref{fig14} \& \ref{fig15}. Winebarger et al. (2002) found persistent redshifts in legs of the cooler 1~MK loops that are seen in TRACE 171~{\AA}, although no such correlation is seen here.

\section{Magnetic Field Extrapolation}

Figure 16 shows TRACE 171~{\AA} images taken at approximately 12:00 UT on each of the five observing days. The spatial resolution of TRACE is greater than that of the EUVI data presented earlier. However, only one wavelength band (171~{\AA}) was observed. On these high-resolution images we can see strong moss emission (indicated by the letters M{\sc s} on Figure 16) on the left-hand side of the AR on Days 1 and 2. We can also see the footpoints of large loop systems (CF) stretching out from the west to the east of the AR, and sweeping around the north and south. On Day 3 we start to see the most notable difference; loops start to brighten in this passband closer to the centre of the AR (L{\sc b}) as was previously shown with EIS data, indicating that the central AR is beginning to cool in agreement with the XRT analysis. Then on Day 4, the bright feature (F{\sc b}) on the right-hand side of the AR, as seen in EIS Fe~{\sc xii} and Fe~{\sc xiii} images, is also present in 171~{\AA} at 1 MK. It appears to be the footpoint of a loop that is cooling into the EUV passbands.


\begin{figure*}
\centering
\figurenum{16}
\plotfiddle{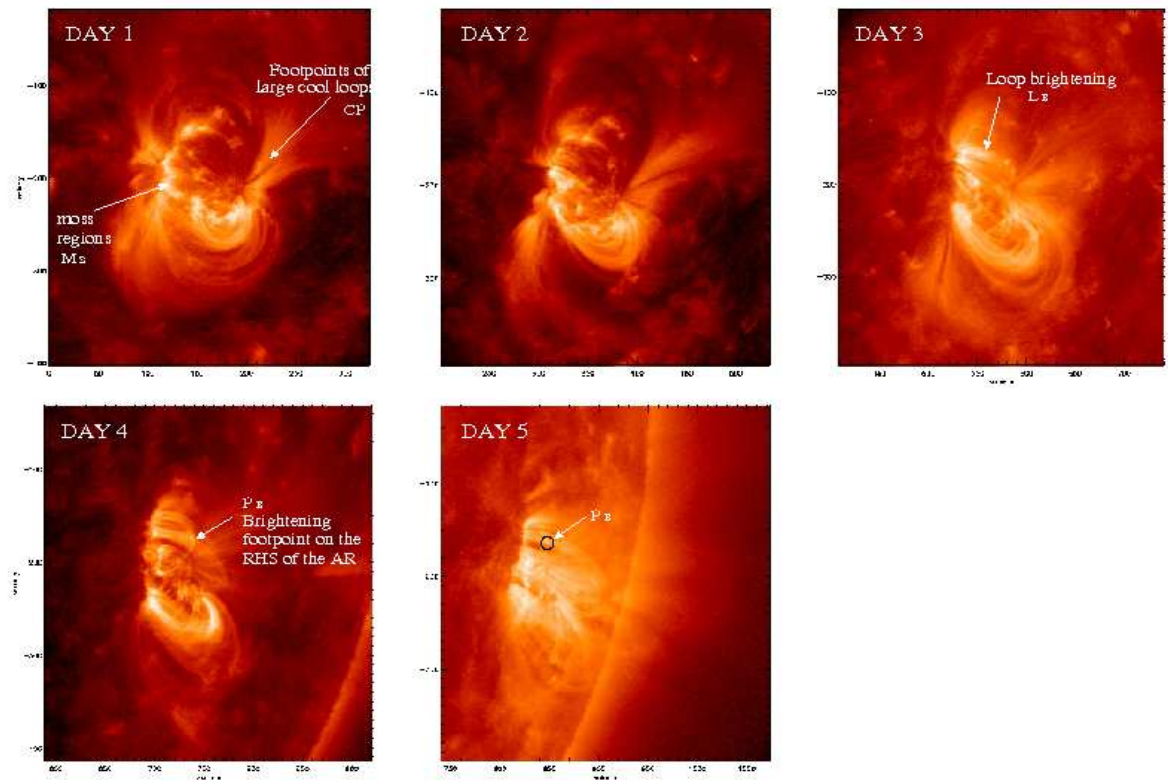}{1.in}{0}{450}{300}{-0}{-0}
\caption{TRACE 171~{\AA} images of the active region taken at around 12:00~UT for every day of the observing run. \label{fig16}}
\end{figure*}


Figure 17 shows a contoured TRACE 171~{\AA} image from Day 1 co-aligned with and overplotted onto a full-disc SOHO/MDI magnetogram image. We see that the bright EUV moss regions match very well with the dispersed positive magnetic field. However, the negative field which consists of one large area of concentrated magnetic flux representing a sunspot, has no such mossy features; generally it appears dark on all EUV and X-ray images. 


To try to understand the morphology of this AR and how the EUV and X-ray emission relates to the magnetic flux observed at the photosphere, we applied the standard technique of extrapolating the magnetic field in the solar atmosphere above the photosphere using the MDI magnetogram data to provide the lower boundary condition. We performed such a simple potential extrapolation for Days 1, 2 and 3. After this time the AR had rotated too far towards the solar limb for the MDI data to be considered accurate. The potential extrapolation was chosen for its simplicity, and because we wanted to make a visual assessment (via a comparison of the extrapolated fieldlines with the X-ray and EUV data) of how far from potential the real AR magnetic field was. 


A magnetic feature on a magnetogram is often defined as a set of adjoining pixels, each of which has an absolute data value greater than some chosen threshold. This selects out regions of strong magnetic field, which may reasonably be expected to determine the large-scale structure of the 3D field. Brian Welsh's feature tracking code (YAFTA; DeForest et al. 2007) was used to track the magnetic features observed on the MDI magnetograms, within a selected area centred on the active region. Each of these magnetic features was then reduced to a point source, with its charge determined by the integrated flux of its parent magnetic feature, and its location given by a flux-weighted average of the pixel locations of the parent feature. This set of point magnetic sources was then fed into Dana Longcope's magnetic topology code (MPOLE; Longcope \& Klapper 2002) which performs the actual magnetic field extrapolation, and calculates the 3D magnetic topology of the resulting field.


\begin{figure}[!h]
\centering
\figurenum{17}
\plotfiddle{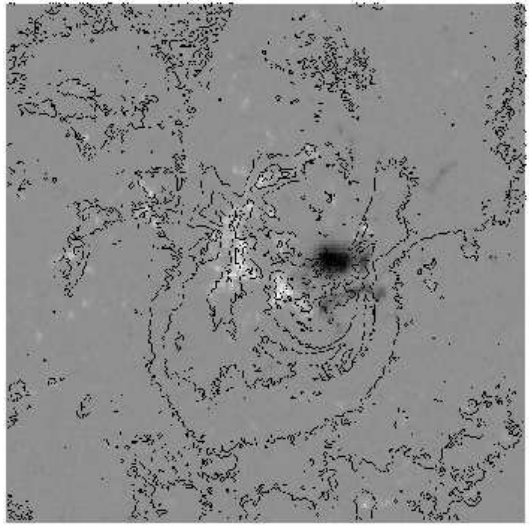}{1.in}{0}{200}{200}{-0}{-0}
\caption{TRACE 171~{\AA} ({\it contours}) over-plotted on a SOHO/MDI magnetogram image of the active region from Day 1. \label{fig17}}
\end{figure}


The code locates all the magnetic null points, along with their associated separatrix surfaces, and spine and separator fieldlines (Longcope 2005). Together, these topological features are referred to as the \emph{topological skeleton} of the magnetic field. Once a set of poles and nulls has been obtained for a given magnetogram, it is possible to map the footprint of the field's skeleton, \emph{i.e.} its intersection with the photospheric plane. This is very helpful in visualising the full topology of the field. Figure 18 shows the skeleton for Days 1, 2, \& 3. The solid lines denote the photospheric spines and the dashed lines represent the intersections of the separatrix surfaces with the photosphere.


\begin{figure*}
\centering
\figurenum{18}
\plotfiddle{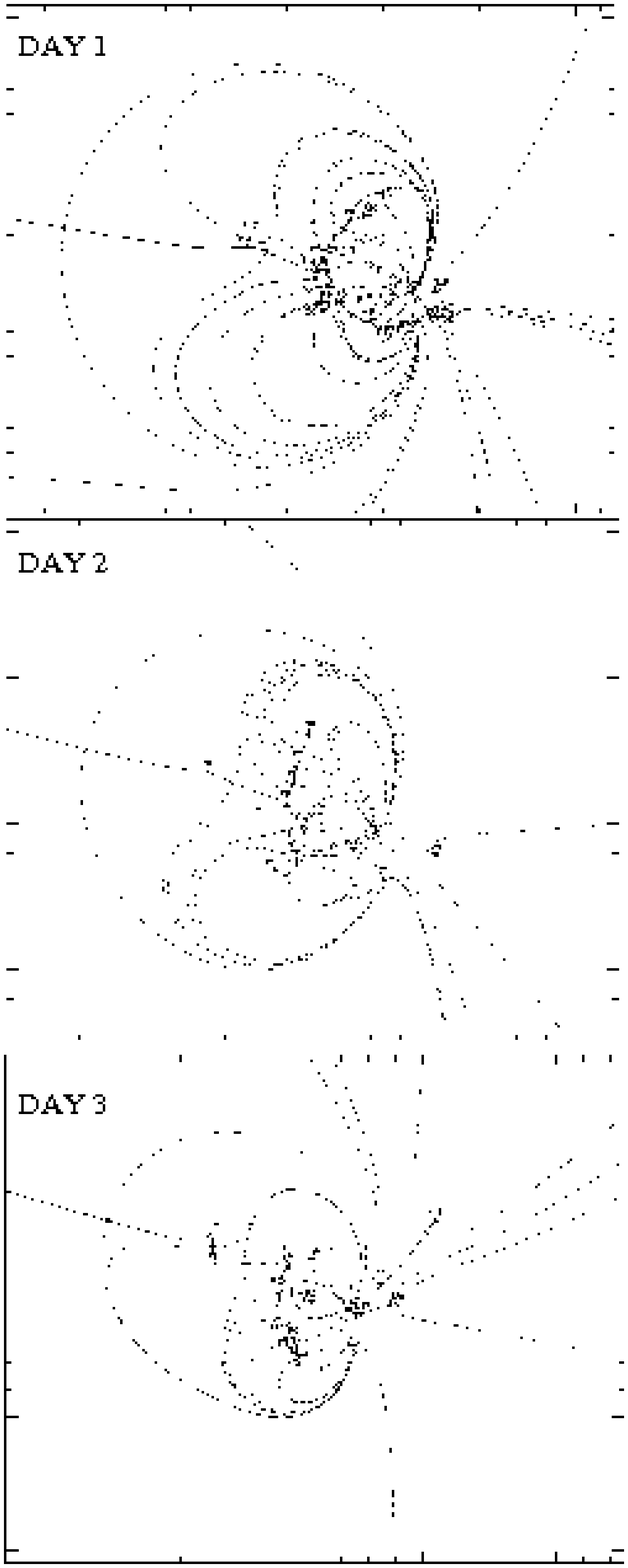}{1.in}{0}{200}{550}{-0}{-0}
\caption{Diagram showing the magnetic poles, the calculated null points and hence the extrapolated footprint of the field's skeleton for Days 1, 2 \& 3. \label{fig18}}
\end{figure*}


Unsurprisingly, the topological structure of the field is dominated by the strong negative sources representing the sunspot. The moss region shows up as a scattering of many relatively weak positive sources forming an arc shape to the left of the negative sources. All of the fieldlines from the moss sources connect to the sunspot. However as the sunspot sources are stronger, fieldlines from the sunspot also leave the box to connect to other distant sources. This is definitely an example of a bipolar (as opposed to quadrupolar) source configuration in an active region.

Using the photospheric footprints as a base, we can choose any two magnetic sources (one positive and one negative) and reconstruct a representative set of the fieldlines that connect them, if indeed they are magnetically connected. Figures 19 \& 20 show examples of such extrapolated fieldlines over-plotted on Ti-poly XRT images and TRACE 171~{\AA} images respectively. Although only a potential magnetic field was used, the fieldlines do seem to follow the X-ray loop emission quite closely in both position and orientation. It can also be seen that where the X-ray emission is brightest there are two sets of loops present; small underlying loops and larger ones above, as is clearly demonstrated in Figure 19.

\begin{figure*}
\centering
\figurenum{19}
\plotfiddle{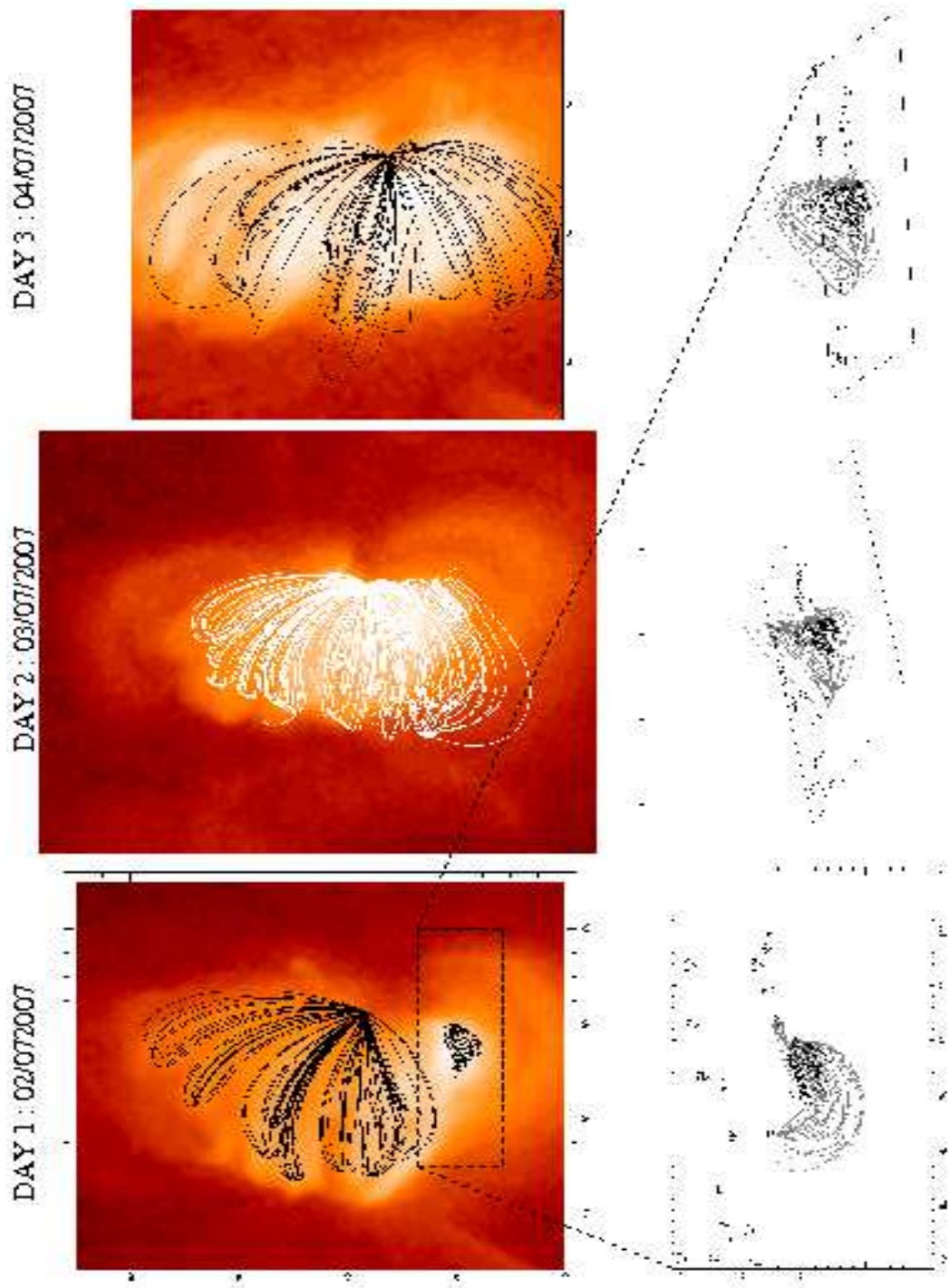}{1.in}{90}{550}{400}{-0}{-0}
\caption{{\it Top:} XRT intensity images for Days 1, 2 \& 3, with individual extrapolated magnetic field lines overlaid from a set negative to a set positive flux point. {\it Bottom:} Close up of the bright XRT region in the north of the AR from Day 1, showing how shorter low lying loops ({\it black}) are present beneath larger taller loops ({\it grey}). \label{fig19}}
\end{figure*}


Most of the fieldlines originate from the large negative flux region on the right-hand side and fan out to the array of smaller positive flux areas on the left-hand side of the AR. This explains the emission seen in the X-rays where loop systems seem to be brightening in every direction from one central point. However, at first glance, there is little direct correlation between the extrapolated fieldlines and the TRACE 171~{\AA} intensities. In fact, the bright EUV moss emission appears to correspond to the left-hand footpoints of the hotter X-ray loops. A similar set of observations was reported by Tripathi et al. (2008). As mentioned previously, there is only moss emission present on the left-hand side of the loop system and not the right-hand side, probably due to the strong negative magnetic field of the sunspot on the right-hand side possibly suppressing the emission from these footpoints. The top diagram from Figure 20 also shows that the large cool loops seen in TRACE are in fact close to potential, following the separatrix surface fieldlines (the dashed lines).

\begin{figure*}
\centering
\figurenum{20}
\plotfiddle{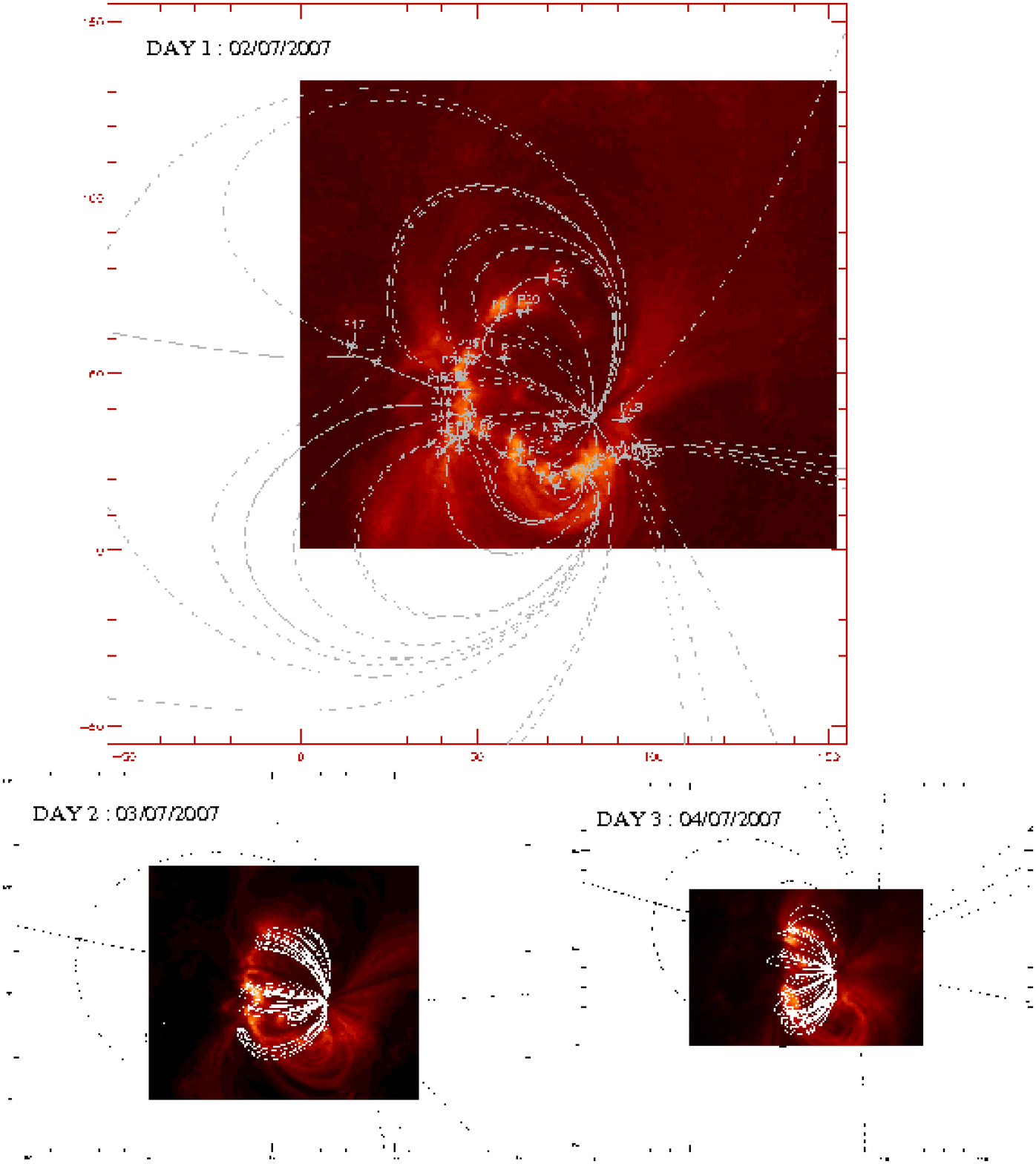}{1.in}{0}{350}{400}{-0}{-0}
\caption{Same as Figure~\ref{fig19} but using TRACE intensity images and showing the broader skeleton of the magnetic field extrapolation. \label{fig20}}
\end{figure*}


\section{Discussion \& Conclusions}

In this paper we have used five different instruments on-board four separate satellite missions to deconstruct the nature of a typical evolving active region over a five day period. The unique thing about this analysis is that after piecing together all of the information we can see more clearly how the EUV, X-ray and magnetogram data fit together in order to extract a global picture of the target.

The filter ratios applied to the EUVI data on Day 1 of the observations appear to produce temperatures from 0.5 - 2.0 MK; however, the 195~{\AA} / 171~{\AA} ratio values did not correspond to any temperatures above 1.1~MK. In contrast the global XRT temperature maps for Days 1 \& 2 showed that most of the emission lay above 2~MK. When we compare the XRT images to the EUVI 171~{\AA} and 195~{\AA} images we can see that there is very little similarity between the observed features.

The XRT temperature maps for Days 3, 4 and 5 only produced temperatures of $\sim$ 1.0 - 2.0~MK. Here, we can start to see cooler temperatures as we are using at least five different passbands to construct the temperature maps, making them more detailed and allowing them to pick up a wider range of temperatures. The AR appears to be cooling over the observational time period as there is little emission seen above 2~MK after Day 2. This can be clearly demonstrated by comparing the EUVI 284~{\AA} images to the X-ray images as the observed structures at these two wavelengths grow in similarity from Day 1 through to Day 5. 

The potential magnetic field extrapolation was sufficient to mimic the X-ray loop regions, in fact, the large cool (171~{\AA}) TRACE loops appear close to potential following the separatrix surface fieldlines. The very bright X-ray intensities observed towards the south of the central AR were shown to be the result of a small set of X-ray loops sitting directly beneath a larger array of loops and not due to high densities or temperatures. 

When looking at individual AR features with the double filter ratio method it was seen that different structures (i.e. moss regions, hot loops and cool loops) appear to occupy different areas of space around the temperature curve. Hence, it maybe possible to distinguish between them thermally to a certain degree. The moss regions lay close to the curve probably due to the fact that they were visible in all three temperature passbands. The cooler regions were situated towards the bottom left-hand corner of the diagram showing a decrease in both the 195~{\AA} / 171~{\AA} and the 284~{\AA} / 195~{\AA} ratios, demonstrating a clear move to cooler temperature passbands as would be expected. This detour from the colour-colour curve could be due to the dynamic cooling of that region. However, it could also be the case that if the count rate is seen to be low on one of the images, in any particular area being studied, then the errors can become very large for that ratio causing the point to move away from the curve. In this case there is a distinct lack of emission seen in the 284~{\AA} passband. 

The exact opposite was shown to be true for the hotter central part of the AR; it is always hotter than the outer regions, for the entire observing period and these points move into the hotter passbands on the colour-colour curve demonstrating an increase in both the ratio values. Equally this could be due to heating within the AR core or the lack of emission seen in the 171~{\AA} passband. 

From the EIS data, the global Fe~{\sc xii} density maps (1.3~MK) showed higher densities (1.0$\times$10$^9$ - 3.0$\times$10$^{10}$~cm$^{-3}$) than Fe~{\sc xiii} (1.6~MK) maps (5.0$\times$10$^8$ - 6.0$\times$10$^9$~cm$^{-3}$), although the high density areas always corresponded to areas of strong intensity emission. This was not the case for the X-ray temperature maps. A good example of this was seen on Day 4 when a bright intensity feature seen in both XRT and EIS appeared dark on the temperature map but bright on the density map, highlighting the fact that brighter emission isn't necessarily hotter emission; in this case it is a localised area of denser plasma. The EIS density maps show high densities corresponding to strong moss emission. The moss corresponds to temperatures between 0.8 - 1.6~MK and is also directly linked to the positive MDI magnetic field.  The velocity maps for this AR showed fairly strong blueshifts above the large negative polarity region and also at boundaries between ``hot'' and ``cool'' emission, whereas the majority of the loops within the central AR displayed strong redshifts (13 - 25~km~s$^{-1}$).

\begin{figure}[!h]
\centering
\figurenum{21}
\plotfiddle{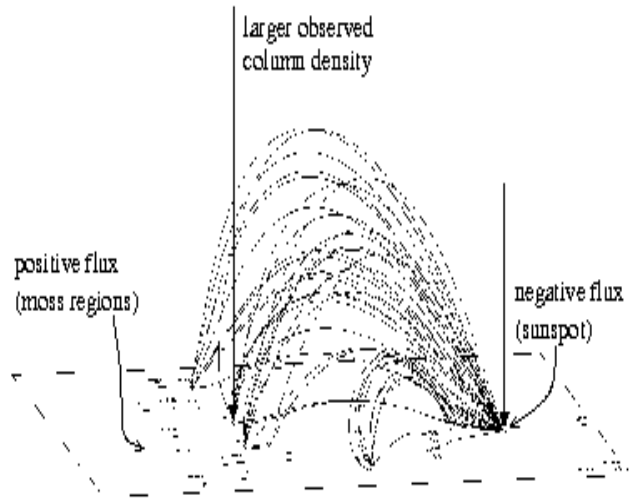}{1.in}{0}{250}{200}{-0}{-0}
\caption{Profile view of the extrapolated magnetic fieldlines taken from Day 1 over the whole active region. \label{fig21}}
\end{figure}

Why do we only see high densities and multi-thermal EUV moss regions over smaller and weaker magnetic field regions and not over the larger stronger magnetic field areas? There is no corresponding EUV or X-ray emission seen above the negative field. Could it be a line-of-sight issue? Taking into account the topology of the magnetic fieldlines seen in Figure~\ref{fig21}, the loops all appear to arch up to a high altitude directly above the positive magnetic field which would create an increasingly large column density and therefore the observed high density moss regions down the left-hand side of the AR. 

Is it to do with the plasma flow? Large, and therefore stronger magnetic flux features (Fludra \& Ireland 2008) such as the negative one seen on these observations are known to inhibit the plasma motions in the photosphere. The positive field on the other hand is seen as many smaller less intense magnetic field regions, which individually are not strong enough to suppress photospheric flows. Therefore, asymmetric heating could be a likely scenario in this situation (Winebarger et al. 2008) and consequently moss regions are only seen on one side of the loop system.

Could this AR be affected/heated by localised magnetic reconnection? Although there are no large-scale changes to the topology over Days 1 to 3, there are some movements of the positive magnetic field sources on the photosphere and a small flux cancellation event on the far right-hand side of the sunspot between Days 1 \& 2. Such driving motions can inject energy into the AR, which can then be released as heat via reconnective processes at null points (Pontin \& Galsgaard 2007) or separatrix surfaces (Priest et al. 2005). Reconnection at null points near the photosphere could be responsible for the EUV moss emission, and separatrix surface reconnection could be responsible for the X-ray loop emission.

In conclusion, this typical, small, evolving AR showed observable changes over the five day period. By using a range of instrumentation and techniques we have been able to understand these physical changes and unite the observed differences seen in the data to produce a broader view of the region. Future work will include analysing a range of active regions with similar instrumental coverage, and it is our intention that HOP~18 will be run again when the Sun exits its current solar minimum.

\acknowledgements
We would like to thank Peter Young for all of his help with the Hinode EIS data reduction and analysis. Hinode is a Japanese mission developed and launched by ISAS / JAXA, collaborating with NAOJ as a domestic partner, NASA and STFC (UK) as international partners. Scientific operation of the Hinode mission is conducted by the Hinode science team organized at ISAS / JAXA. This team mainly consists of scientists from institutes in the partner countries. Support for the post-launch operation is provided by JAXA and NAOJ (Japan), STFC (UK), NASA, ESA, and NSC (Norway). CHIANTI is a collaborative project involving the NRL (USA), RAL (UK), and the following Universities: College London (UK), Cambridge (UK), George Mason (USA), and Florence (Italy). TRACE and SOHO are projects of international cooperation between ESA and NASA. The STEREO/SECCHI data used here are produced by an international consortium of the Naval Research Laboratory (USA), Lockheed Martin Solar and Astrophysics Lab (USA), NASA Goddard Space Flight Centre (USA), Rutherford Appleton Laboratory (UK), University of Birmingham (UK), Max-Planck-Institut f\"{u}r Sonnensystemforschung (Germany), Centre Spatiale de Li\'{e}ge (Belgium), Institut d'Optique Th\'{e}orique et Appliqu\'{e}e (France), and Institute d'Astrophysique Spatiale (France).

\end{document}